\documentclass[twocolumn,pra,superscriptaddress,nofootinbib]{revtex4} 
\pdfoutput=1
\usepackage{setspace}
\usepackage{amsfonts,amsmath,amssymb}
\usepackage{txfonts}
\usepackage{graphicx, color}
\usepackage{verbatim}
\usepackage{hyperref}
\usepackage{bm}

\begin{document}

\title{The Beliaev Broken Symmetry Description of Superfluidity vs the Classical-Field Approach} 

\author{Allan Griffin}
\thanks{Allan Griffin passed away on 19 May 2011 after writing the first draft of this manuscript.}
\affiliation{Department of Physics, University of Toronto, 60 St.~George Street, Toronto, ON M5S 1A7, Canada} 
\author{Eugene Zaremba}
\affiliation{Department of Physics, Engineering Physics and Astronomy, Queen's University, Kingston, ON K7L 3N6, Canada} 

\begin{abstract}
\begin{center}
`The Lord of the rings, the One Ring to rule them all'\\ 
from J.~R.~R. Tolkien \cite{tolkien_book_54}
\end{center}
The standard theoretical basis for understanding superfluidity in Bose systems was formulated by Beliaev in 1957, based on splitting the quantum field operator into a macroscopically occupied condensate component and a non-condensate component.  This leads to a description of the condensate in terms of a `single-particle state', the so-called macroscopic wavefunction.  Since the discovery of Bose-condensed gases, an alternative theoretical picture has been developed which is based on a `coherent band' of classically occupied states.  This is often called the classical or c-field approach.  The goal of this chapter is to review the differences between the Beliaev broken symmetry and c-field approach, and to argue that the c-field concept of a coherent condensate band of states has problems as a description of Bose superfluidity. However, the c-field idea of treating the lowest energy excitations classically can be used to advantage to simplify calculations within the Beliaev broken-symmetry formalism. 
\end{abstract}

\maketitle

\section{Introduction}
\vspace{-0.5\baselineskip}
One of the major goals of the Sandbjerg and the Durham workshops was to bring together theorists who were using different methods for dealing with Bose gases at finite temperatures and in non-equilibrium states.  The hope was that this would lead to a critical appraisal of the different formalisms, possibly including `benchmark' problems where the results could be compared. These goals have been partially met.  In particular, Proukakis and Jackson~\cite{proukakis_jackson_08} have published a very useful article comparing many different formalisms for treating Bose superfluids at finite temperatures.  However a major problem in the recent literature is that papers based on the ZNG formalism~\cite{zaremba_nikuni_99,griffin_nikuni_book_09} and the classical field approach~\cite{brewczyk_gajda_07,blakie_bradley_08,cockburn_proukakis_09} are like `two ships passing in the night'.  Apart from a few references, papers using one approach largely ignore results using the other approach. A major review article~\cite{blakie_bradley_08} on the c-field approach has 223 references, but the key paper setting up the ZNG formalism~\cite{zaremba_nikuni_99} is mentioned in one paragraph.  By the same token, the recent monograph~\cite{griffin_nikuni_book_09} by Griffin, Nikuni and Zaremba (henceforth referred to as GNZ) only makes brief contact with the final results of c-field theories in Chapters 8 and 9 when deriving the simple dissipative Gross--Pitaevskii Equation (DGPE).

In an attempt to remedy this situation and gain greater understanding, we initiate in this chapter a critical discussion of the conceptual basis of the classical field approach~\cite{proukakis_jackson_08,blakie_bradley_08,brewczyk_gajda_07,stoof_99}, comparing it with the Beliaev broken-symmetry theory of superfluids (see Ref.~\cite{beliaev_58} and also Chapters~4--5 of GNZ).  The latter remains the basis of almost all current discussions of superfluidity in interacting Bose systems in the literature using quantum field theoretic techniques (see, for example, Ref.~\cite{capogrosso_giorgini_10}).  The ZNG coupled equations [Eqs.~(3.21) and (3.42) of GNZ] are based on the Beliaev scenario, albeit with the addition of various approximations (`Popov' approximation for the neglect of the pair anomalous average~\cite{griffin_96}, Hartree--Fock spectrum for the  excitations, etc.).  However, as discussed in Chapters 6 and 7 of GNZ, the simplified ZNG formalism can be extended in a systematic manner within the Beliaev field-theoretic approach.

We hope our comparison of the basis of the c-field approach with the Beliaev formulation will make a useful contribution to the current literature.  For references and more detailed discussion of the topics covered in this chapter, we refer to recent reviews~\cite{proukakis_jackson_08,blakie_bradley_08,griffin_nikuni_book_09,brewczyk_gajda_07}.  An alternative viewpoint on this important comparison is given in Ref.~\cite{wright_davis_chapter_12}.

\vspace{-0.3\baselineskip}
\section{Methodology}
\vspace{-0.4\baselineskip}
\subsection{Beliaev Theory of Superfluidity \label{griffin:beliaev}}
\vspace{-0.3\baselineskip}
The key feature of any theory of a Bose-condensed quantum fluid is that it should explain the appearance of a new degree of freedom which describes the phenomenon of superfluidity.  These features were first captured in the Landau two-fluid theory (developed in 1941 for superfluid $^4$He) and were later given a full microscopic basis by Beliaev in 1957~\cite{beliaev_58}.  Beliaev, extending the pioneering work of London~\cite{london_38} in 1938 and Bogoliubov~\cite{bogoliubov_47} in 1947 on dilute Bose gases, formulated a systematic way of separating out the superfluid degree of freedom associated with the appearance of a Bose condensate (BEC) in a quantum field-theoretic description of an interacting system of Bose particles.  In Beliaev's formulation, which is the canonical approach to superfluidity in Bose systems, the superfluid degree of freedom has its microscopic origin in the broken-symmetry average value of the quantum field operator $\phi(\mathbf{r})=\langle{\hat\Psi}(\mathbf{r})\rangle_{\rm BS}$. This can be viewed as a single-particle state which is macroscopically occupied and plays the role of the order parameter of the new superfluid phase appearing below some transition temperature $T_{\mathrm{c}}$.  The concept can also be extended to non-equilibrium situations where the ensemble average is taken with respect to a non-equilibrium density matrix.

The Beliaev formulation~\cite{beliaev_58} is based on a decomposition of the quantum field operator ${\hat\Psi}(\mathbf{r})$ into a condensate component $\phi(\mathbf{r})$ as defined above plus a non-condensate component\footnote{Note that the ZNG discussion \cite{griffin_nikuni_book_09} uses $\Phi$ rather than $\phi$ for the condensate and $\tilde\psi$ rather than $\hat \psi'$ for the non-condensate fluctuations.},
\begin{equation} {\hat\Psi}(\mathbf{r}) = 
\phi(\mathbf{r})+{{\hat\psi}'}(\mathbf{r}).\label{belsuper1}
\end{equation}
Invoking this idea, one can then develop a many-body perturbation scheme for the non-condensate fluctuation ${{\hat\psi}'}(\mathbf{r})$ in the presence of the Bose-condensate described by $\phi(\mathbf{r})$.  This takes into account fully the complex correlations between $\phi(\mathbf{r})$ and ${{\hat\psi}'}(\mathbf{r})$.  Because it describes a single-particle state which is macroscopically occupied, in most treatments $\phi(\mathbf{r})$ is treated as a classical field.  However this latter approximation is not essential to the Beliaev broken-symmetry formalism~\cite{talbot_griffin_84,talbot_thesis_83}.

A crucial aspect of the Beliaev formulation of superfluidity is that the macroscopic wavefunction $\phi(\mathbf{r})$ can be complex 
\begin{equation}\phi(\mathbf{r}) = 
{\sqrt{n_{\mathrm{c}}(\mathbf{r})}}e^{i\theta(\mathbf{r})},\label{belsuper2}
\end{equation}
where $n_{\mathrm{c}}(\mathbf{r}) = |\phi(\mathbf{r})|^2$ defines the condensate density. The phase of this order parameter naturally leads to an associated condensate velocity field
\begin{equation}\mathbf{v}_{\mathrm{c}}(\mathbf{r}) = 
\frac{\hbar}{m}{\mathbf{\nabla}}\theta(\mathbf{r}).\label{belsuper3}
\end{equation}
Clearly $\mathbf{v}_{\mathrm{c}}(\mathbf{r})$ is irrotational, a key element of the Landau theory of superfluidity. Choosing a finite value of the order parameter $\phi(\mathbf{r})$ corresponds to a specific choice of the phase, which breaks the gauge symmetry. This is most conveniently done by using a number non-conserving formalism~\cite{griffin_nikuni_book_09,beliaev_58,talbot_griffin_84,talbot_thesis_83,griffin_book_93}.

Using linear response theory to describe the effect of a condensate  moving with a velocity $\mathbf{v}_{\mathrm{c}}$, one can calculate how it drags the non-condensate atoms along. The total superfluid mass current can be rigorously proved [see Eq.~(3.3) of Ref.~\cite{talbot_griffin_84}] to be given by ${\bf j}_\mathrm{s} = \rho_\mathrm{s}\mathbf{v}_{\mathrm{c}}$, where the superfluid density $\rho_\mathrm{s}(\mathbf{r})$ is fundamentally distinct from $n_{\mathrm{c}}(\mathbf{r})$ and is given as the difference between the longitudinal and transverse velocity response functions (see pages~128--131 of Ref.~\cite{griffin_book_93}). Thus the Beliaev broken-symmetry formalism based on Eqs.~(\ref{belsuper1}) and (\ref{belsuper2}) naturally leads to two fundamental features of a Bose superfluid, namely (a) the superfluid velocity $\mathbf{v}_\mathrm{s}$ is equal to the condensate velocity $\mathbf{v}_{\mathrm{c}}$, and (b) a rigorous microscopic definition for the superfluid density, which reduces to Landau's quasiparticle expression in two-fluid hydrodynamics.  These two features do not have to be inserted into the theory as additional assumptions.

The Beliaev picture, based on Eq.~(\ref{belsuper1}), is the conceptual and computational basis of most microscopic studies of superfluidity in liquid $^4$He.  The work of ZNG~\cite{griffin_nikuni_book_09} as applied to trapped Bose gases is based on this approach, albeit within a simplified treatment of the non-condensate fluctuations. One of the main strengths of the Beliaev approach is that it provides a microscopic basis for the derivation of Landau's phenomenological two-fluid hydrodynamics which is valid when interactions are strong enough to establish local thermodynamic equilibrium.  Likewise, it can be used to derive Landau--Khalatnikov hydrodynamics (see Chapter~17 of GNZ) which accounts for the damping that is associated with dissipative transport coefficients.  In this derivation, a natural definition of what we `mean' by the superfluid density $\rho_\mathrm{s}(\mathbf{r})$ emerges, based on the underlying Bose condensate density $n_{\mathrm{c}}(\mathbf{r})$.  While $\rho_\mathrm{s}(\mathbf{r})$ can be quite different from $n_{\mathrm{c}}(\mathbf{r})$, as noted above, the superfluid velocity field $\mathbf{v}_\mathrm{s}(\mathbf{r})$ is equal to the condensate velocity field $\mathbf{v}_{\mathrm{c}}(\mathbf{r})$ defined in Eq.~(\ref{belsuper3}). Being able to derive Landau's two-fluid equations under appropriate conditions (short collision times) is an essential requirement of any complete microscopic theory of superfluidity.

It is clear that in the Beliaev formulation, there is a unique single-particle state described by $\phi(\mathbf{r})$ which has a `privileged status' among other single-particle states and describes the new superfluid phase.  To quote J.~R.~R. Tolkien when he refers to the 19 rings produced and distributed by Sauron of Mordor~\cite{tolkien_book_54}, one might say that $\phi(\mathbf{r})$ is `The Lord of the rings, the One Ring to rule them all'.  In recent years, however, a seemingly different picture of describing interacting Bose gases has emerged~\cite{proukakis_jackson_08,blakie_bradley_08,brewczyk_gajda_07,stoof_99} in which $\phi(\mathbf{r})$ has no special status.  This approach is now often called the `classical-field approach', or more succinctly, the c-field approach.  In Section~\ref{griffin:c_field} we review the conceptual basis of this approach.  In Sections~\ref{griffin:conceptual} and \ref{griffin:excitations}, we then argue that despite its success as a computational method, it has problems when it comes to addressing superfluidity in interacting Bose systems.

\subsection{C-Field Approach \label{griffin:c_field}}
Recently there has been considerable work on Bose gases at finite temperatures using what is called the c-field approach (see Refs.~\cite{davis_wright_chapter_12} and~\cite{cockburn_proukakis_chapter_12} for a general overview). The key idea behind this approach is based on the observation that the lowest energy eigenstates of a trapped Bose gas have occupation numbers much larger than unity and hence can be treated classically~\cite{kagan_svistunov_92a,kagan_svistunov_92b,kagan_svistunov_94,kagan_svistunov_97}. One variant of this approach~\cite{blakie_bradley_08} divides the states of an interacting Bose gas into two classes: a `coherent' band of states which is treated by means of a classical field, and an `incoherent' band of higher energy states which must be treated in some other way. The classical field can be introduced systematically by means of a projection technique (as discussed further below).  The c-field approach is an attractive numerical scheme since the states in the coherent band satisfy an equation analogous to the usual Gross--Pitaevskii Equation (GPE) valid at $T=0$. 
As noted above, a very detailed review article~\cite{blakie_bradley_08} has been published discussing the c-field approach. It sets up the conceptual basis for this approach, with extensive references to earlier work, and shows how various kinds of projected and stochastic Gross--Pitaevskii-type equations can be derived. The projected Gross--Pitaevskii equation (PGPE) omits entirely the coupling to the incoherent region~\cite{davis_morgan_02}, whereas the so-called stochastic theories include it via dissipation and stochastic noise terms.  These latter theories include the closely-related stochastic projected GPE (SPGPE) of Gardiner {\em et al.}~\cite{gardiner_davis_03} and the stochastic GPE (SGPE) approach of Stoof~\cite{stoof_bijlsma_01}.  In the following, we focus on the conceptual issues underlying the PGPE approach (see Refs.~\cite{brewczyk_gajda_07}~and~\cite{blakie_bradley_08} for details regarding the numerical implementation of the c-field scenario), although many of our comments are also relevant to the SPGPE and SGPE approaches.  We refer separately to the SGPE approach of Stoof in Section~\ref{griffin:stoof}.

In Section~\ref{griffin:beliaev}, we pointed out that the Beliaev formulation for dealing with a Bose system of particles with a condensate is based on the separation of the quantum field operator $\hat\Psi(\mathbf{r})$ into two distinct parts, as given in Eq.~(\ref{belsuper1}).  The end result is a rigorous scheme for treating the dynamics of the non-condensate field operator ${\hat\psi}^\prime(\mathbf{r})$  which is dynamically coupled to the superfluid degree of freedom $\phi(\mathbf{r})$.  This operator of course also describes the non-condensate in the normal phase.  The c-field approach adopts a somewhat similar point of view, except that the quantum field operator ${\hat{\Psi}}(\mathbf{r})$ is now decomposed in a manner different from Beliaev, namely
\begin{equation} \hat\Psi(\mathbf{r}) 
\equiv\hat \Psi_{\mathbf C}(\mathbf{r}) +\hat\Psi_{\mathbf
I}(\mathbf{r}).\label{belsuper4}\end{equation}
Here the part of $\hat\Psi$ which is referred to as the coherent part $\hat \Psi_{\mathbf C}$ corresponds to a band of single-particle states which can be treated classically since their occupation numbers are assumed to be much larger than unity.  The incoherent part ${\hat\Psi}_{\mathbf I}$ describes the higher energy states which must be treated quantum mechanically.  The best choice of `energy cutoff' $E_{\rm cut}$ which separates these two bands is a delicate question discussed at length in the implementation of the c-field approach~\cite{proukakis_jackson_08,blakie_bradley_08,brewczyk_gajda_07}, but one might expect $E_{\rm cut} \sim k_{\mathrm{B}}T$.

The simplest (and earliest~\cite{kagan_svistunov_92a,kagan_svistunov_92b,kagan_svistunov_94,kagan_svistunov_97}) version of the c-field approach is to restrict oneself to the coherent `condensate' band. The exact Heisenberg equation of motion for $\hat\Psi(\mathbf{r})$ is given by
\begin{equation} i\hbar \frac{\partial\hat\Psi(\mathbf{r}, t)}{\partial 
t} = 
\left[-\frac{\hbar^2\nabla^2}{2m}+V_{\rm 
ext}(\mathbf{r})\right]\hat\Psi(\mathbf{r}, t) + 
g\hat\Psi^\dag(\mathbf{r}, 
t)\hat\Psi(\mathbf{r}, t)\hat\Psi(\mathbf{r}, 
t),\label{belsuper5}\end{equation}
where we assume a simple $s$-wave delta function pseudopotential of strength $g$. Writing $\hat\Psi(\mathbf{r})$ as an expansion in an appropriate basis of single-particle eigenstates $\varphi_i(\mathbf{r}),$ 
\begin{equation} \hat\Psi(\mathbf{r}) =\sum_i
\varphi_i(\mathbf{r})\hat a_i,\label{belsuper6}\end{equation}
the coherent part in Eq.~(\ref{belsuper4}) can be expressed as the projection onto the states comprising the coherent band
\begin{eqnarray} {\hat\Psi}_{\mathbf C}(\mathbf{r}) &=& {\cal{P}}_{\mathbf 
C}\left\{\hat\Psi(\mathbf{r})\right\}\nonumber\\
&\equiv&\sum_{i\in\mathbf{C}}\varphi_i(\mathbf{r})\int 
\!d\mathbf{r}'\varphi^\ast_i(\mathbf{r}')\hat\Psi(\mathbf{r}').\label{bels
uper7}\end{eqnarray}
To the extent that this coherent part of the full quantum field operator describes states which can be treated classically [i.e., ${\hat\Psi}_{\mathbf C}(\mathbf{r})$ becomes a c-number function], it satisfies the equation of motion  [cf. Eq.~(\ref{belsuper5})]
\begin{eqnarray} i\hbar \frac{\partial\Psi_{\mathbf 
C}(\mathbf{r},t)}{\partial t} 
&=&\left[-\frac{\hbar^2\nabla^2} {2m}+V_{\rm 
ext}(\mathbf{r})\right]\Psi_{\mathbf C}(\mathbf{r},t)\nonumber\\
&+&{\cal{P}}_{\mathbf C}\left\{g|\Psi_{\mathbf C}(\mathbf{r},t)|^2\Psi_{\mathbf 
C}(\mathbf{r},t)\right\}.\label{belsuper8}\end{eqnarray}
This projected Gross--Pitaevskii equation (PGPE) is quite different from the usual GPE since $\Psi_{\mathbf C}(\mathbf{r},t)$ describes a band of states. However, the techniques for solving Eq.~(\ref{belsuper8}) are very similar to those used to solve the GPE for $\phi(\mathbf{r},t)$ at $T=0$.

An essential difference between the c-field approach and the Beliaev formulation is already apparent.  $\Psi_{\mathbf C}(\mathbf{r})$ describes a `band' of low energy single particle eigenstates, rather than the lowest one $\phi(\mathbf{r})$.  This band is sometimes referred to as a `coherent' or `condensate' band of states and plays a crucial conceptual role in c-field treatments.  Equally important, it forms the basis for numerical approximations~\cite{proukakis_jackson_08,blakie_bradley_08,brewczyk_gajda_07}. It is correctly argued that this condensate band (even when the incoherent part is ignored) contains an essential aspect of the physics of Bose superfluids at finite temperatures which is omitted when only the average $\phi(\mathbf{r})$ is considered, namely the fluctuations of the condensate.

\section{Validity}

\subsection{Conceptual Problems with the C-Field Approach\label{griffin:conceptual}} 
In our view, a major shortcoming of the c-field theories is that no fundamental distinction is made between the different low energy states contributing to the coherent band.  Indeed, the classically occupied states comprising the coherent band are sometimes viewed as defining a `kind of order parameter' which generalises the single state $\phi(\mathbf{r})$ used in the Beliaev broken-symmetry theory (see, for example, Refs.~\cite{stoof_99,proukakis_jackson_08}).  In such treatments, it is not made clear how this band of states can be considered an `order parameter' in the same way that $\phi(\mathbf{r})$ in the symmetry-breaking scenario is associated with a new superfluid phase with off-diagonal long-range order (ODLRO).  It simply picks out a class of low energy states which can be treated classically.  This set of states is certainly not the `condensate' in the conventional sense (as also highlighted in those works).  In order to make contact with the condensate in the Bose broken-symmetry picture, the c-field methods invoke a Penrose--Onsager analysis~\cite{penrose_onsager_56,pitaevskii_stringari_03} to diagonalise the single-particle density matrix  $\rho^{(1)}(\mathbf{r},\mathbf{r}',t)=\langle\hat\Psi^\dag(\mathbf{r},t)\hat\Psi(\mathbf{r}',t)\rangle$ where the average is to be taken with respect to a non-equilibrium density matrix.  In the c-field (PGPE) approach, $\hat\Psi(\mathbf{r},t)$ is replaced by $\Psi_{\mathbf C}(\mathbf{r},t)$ and the average is evaluated as a time average.  The lowest eigenstate is then taken to define the Bose condensate.  This approach is clearly reasonable for obtaining the condensate in the {\it equilibrium} state but some different method must be used to make contact with the time-dependent order parameter $\phi(\mathbf{r},t)$ that would appear in a non-equilibrium situation.  The latter quantity appears directly when Bose broken-symmetry is invoked in non-equilibrium theories such as the ZNG theory.  However, it should be noted that in the context of the SGPE, $\phi(\mathbf{r},t)$ can also be extracted by performing an additional ensemble average over a distribution of initial states consistent with the non-equilibrium evolution of interest (see Section~\ref{griffin:stoof}).

As emphasised in our review of the Beliaev formulation in Section~\ref{griffin:beliaev}, the gradient of the phase of $\phi(\mathbf{r})$ introduces a unique superfluid velocity field $\mathbf{v}_\mathrm{s}(\mathbf{r})$. The existence of such a velocity (sometimes viewed as a `phase locking') defines what we mean by superfluidity and is the reason Beliaev's formulation is useful in the microscopic derivation of the two-fluid behaviour of Bose-condensed systems.  In contrast, each of the states arising from the Penrose--Onsager diagonalisation of the `coherent band' would in general be expected to have a different velocity field, related to the phase of each eigenstate.  It of course would be natural to identify the superfluid velocity with the macroscopically occupied state, that is, the one corresponding to the lowest Penrose--Onsager eigenstate.  However, in recent expositions~\cite{proukakis_jackson_08,blakie_bradley_08,brewczyk_gajda_07} of c-field theories there is little discussion of the superfluid velocity field, a concept central to any satisfactory theory of a Bose superfluid.  Whenever the superfluid velocity is addressed there necessarily is a `Lord of the rings'.

\subsection{Excitations in the C-Field and Beliaev Approaches\label{griffin:excitations}} 

It should be noted that the power of Beliaev's `extraction' of the condensate part of $\phi(\mathbf{r})$ as defined in Eq.~(\ref{belsuper1}) lies in the fact that one is able to set up a systematic diagrammatic theory for the non-condensate dynamics completely analogous to that used for a Bose gas in the normal phase~\cite{beliaev_58,talbot_thesis_83,griffin_book_93}. It is far from obvious that a similar perturbation scheme for the incoherent band $\hat{\Psi}_{\mathbf I}(\mathbf{r})$ can be developed based on the c-field decomposition given in Eq.~(\ref{belsuper4}).

Reading the literature on Bose superfluids, one can easily get the wrong impression that the excitation spectrum of the condensate and non-condensate components [obtained via Eq.~(\ref{belsuper1})] are different.  In fact, one of the great triumphs of the field-theoretic analysis of the structure of correlation functions is that it shows explicitly how a condensate couples and hybridises the excitations of the condensate microscopic wavefunction $\phi(\mathbf{r}, t)$ with those of the non-condensate described by the single-particle Beliaev Green's functions for $\hat\psi'(\mathbf{r})$.  In the field-theoretic formalism (see pages~69--74 of GNZ), one has for example the decomposition
\begin{eqnarray} 
G_1(1, 1') &\equiv& -i\langle 
T\hat\psi(1)\hat\psi^\dag(1')\rangle\nonumber\\
&=& -i\langle\hat\psi'(1){\hat{\psi'}}^\dag(1')\rangle 
+{\sqrt{-i}}\phi(1){\sqrt{-i}}\phi^\ast(1')\nonumber\\
&\equiv&\tilde G_1(1, 1')+G_{1/2}(1)G^\ast_{1/2}(1').\label{belsuper9}
\end{eqnarray}
Bose broken-symmetry leads to equations of motion for $\tilde G_1$ and $G_{1/2}$ which are coupled and this in turn leads to both of these functions sharing the {\it same} hybridised excitation spectrum.  In an analogous manner, the single-particle Green's function $\tilde G_1(1, 1')$ and the density response function $\chi_{nn}(1, 1')$ share the {\it same} excitation spectrum (see pages~96--98 of GNZ), a key experimental signature of Bose superfluids.  This common excitation spectrum (induced by the condensate) is implicitly contained in the pioneering Green's function analysis of Hohenberg and Martin (see Section VI of Ref.~\cite{hohenberg_martin_65}) and was exhibited more explicitly at finite temperatures by Cheung and Griffin~\cite{cheung_griffin_71}. A more general way of showing this shared excitation spectrum is to use the diagrammatic dielectric formalism, reviewed in Chapter~5 of Ref.~\cite{griffin_book_93}. For a model calculation showing how the hybridising effect of the condensate couples the excitations of the condensate and non-condensate, see Section 5.4 of GNZ.

This shared excitation spectrum is usually hidden in most theoretical papers since $G_{1/2}, \tilde G_1$, and $\chi_{nn}$ are computed at {\it different} levels of approximation.  As a result, the excitations associated with these fluctuations appear to be different.  In a consistent calculation, as discussed above, all these correlation functions exhibit the {\it same} hybridised excitation spectrum --- the characteristic signature of a Bose superfluid.  To be explicit, the GPE-like equation of motion for the condensate $G_{1/2}$ (i.e., $\phi$) will have the same excitation spectrum as the non-condensate correlation function $\tilde G_1$, but only if we take care to use consistent approximations in both equations of motion.

The preceding discussion indicates that the c-field division of the excitations of a Bose superfluid into two regions, the coherent and incoherent bands, is both artificial and misleading.  In the Beliaev formulation, there is a single excitation branch describing {\it both} the condensate and non-condensate fluctuations.  At low momentum, these excitations in a Bose superfluid always have the characteristic spectrum of Goldstone--Nambu phonons.  At high momentum, the excitations are particle-like in a dilute Bose gas (in contrast to rotons in superfluid $^4$He) and can be described using a Hartree--Fock (HF) spectrum.  Which momentum region has the greatest weight in thermodynamic quantities depends on whether one is considering the very low (phonon-like excitations) or high (particle-like excitations) temperature region of the superfluid phase.

In this regard, the ZNG model calculations carried out to date treat the fluctuations of ${{\hat\psi}'}(\mathbf{r})$ in terms of HF-like excitations moving in a time-dependent self-consistent field produced by the condensate and non-condensate components.  Although the condensate degree of freedom and the thermal cloud of atoms are treated using quite different approximations, the ZNG theory nevertheless gives an excellent account of the frequencies and damping of collective modes over the range of temperatures studied experimentally (see Chapters~11 and 12 of GNZ for a review of applications).  The essential reason for this is that the dynamics of the condensate and non-condensate are {\it coupled} in a physically realistic way.  As discussed in Chapter~7 of GNZ, one can generalise the ZNG equations of motion by treating the thermal cloud atoms in terms of Goldstone--Nambu phonons, thereby extending the validity of the theory to lower temperatures and restoring the constraints on the excitation spectrum imposed by Bose broken symmetry.  However, from a practical point of view, such an extension would have little effect on the calculated damping rates at higher temperatures.

In the review article~\cite{proukakis_jackson_08} comparing different methods for dealing with Bose gases at finite temperatures, it is suggested that the Beliaev formulation (on which ZNG is based) rests on an `artificial' separation into condensate and non-condensate contributions (the latter being the thermal cloud atoms in ZNG).  We maintain that this separation is not artificial but gives a natural way of capturing the two-fluid nature of superfluids resulting from an underlying Bose condensate.  In contrast, as discussed here and in Section~\ref{griffin:conceptual}, the separation of the states into coherent and incoherent bands in the c-field approach is somewhat {\it ad hoc}.  It leads neither to an understanding of the special role of the lowest energy condensate mode and the superfluid velocity field associated with it, nor of the special nature of excitations in a Bose-condensed fluid.

A further limitation of the c-field theories when including the coupling to the incoherent band of high energy states (i.e., in the context of the SPGPE) is that the latter is approximated as an effective heat bath in all (numerical) treatments to date, while the 
microscopic dynamics of the coherent band is worked out in detail.  Without a proper treatment of the dynamic coupling between the coherent and incoherent bands it will be impossible to describe the damping of collective modes in an accurate way. As stated earlier, the natural way to include this coupling is by means of a many-body perturbation theory approach based on Bose broken symmetry.

\subsection{Computational Advantages of the C-Field Approach}

Despite our critical remarks regarding the conceptual basis of the c-field approach, we must concede that being able to treat the low-lying states classically is an important computational advance.  As a result, one can use a PGPE [such as Eq.~(\ref{belsuper8})] to address some non-trivial non-equilibrium problems without having to introduce a lot of formal `machinery' typical of field theoretic calculations based on the Beliaev formalism (diagrammatic perturbation theory, single-particle Green's functions, etc.).  By its nature, the c-field method is able to provide information regarding fluctuations not readily available using other approaches.  In addition, to the extent that the incoherent band can be treated as a thermal bath as done in the SPGPE, it can provide a qualitative understanding of various equilibration processes such as vortex nucleation and relaxation in rotating condensed gases.  It would nevertheless be useful to have more detailed comparisons with other approaches for situations where the dynamics of the incoherent band cannot be neglected.

Another interesting question is to what extent the c-field approach can provide an adequate description of the critical region near $T_{\mathrm{c}}$. It is well known from the theory of second order phase transitions (arising from a broken-symmetry order parameter) that fluctuations not included in a mean field approximation are crucial.  This topic is surprisingly difficult and requires a careful treatment of the infrared divergences that arise when using quantum field theoretic methods (see, for example Refs.~\cite{capogrosso_giorgini_10,holzmann_fuchs_04}).  In the c-field literature, it has been argued that the PGPE for the coherent band naturally includes the low energy critical fluctuations near $T_\mathrm{c}$ (see, for example, Refs.~\cite{blakie_bradley_08,davis_blakie_06}). At some level, this conjecture is reasonable.  However, one would like more analytical studies to determine if such PGPE calculations (which have the virtue of simplicity) indeed capture the results found in the more standard many-body literature~\cite{capogrosso_giorgini_10,holzmann_fuchs_04}.

In summary, we believe the c-field approach as currently implemented is best viewed as a numerical strategy to simplify calculations rather than as an `alternative' theory to the Beliaev approach.

\subsection{Stoof Formalism: Some Brief Remarks \label{griffin:stoof}}
Apart from ZNG and the c-field approach, the only other general scheme which attempts to deal with the non-equilibrium properties of a Bose-condensed gas is the path-integral formalism developed by Stoof~\cite{stoof_99,duine_stoof_01}.  The final equations he derives are closely related to those of the ZNG approach.  However, by formulating a theory for the probability distribution of a bosonic field, fluctuations of the condensate are included. A simplified version of his theory results in a stochastic GPE (SGPE) with a noise term~\cite{stoof_bijlsma_01,duine_stoof_01}.  The presence of the noise term distinguishes this SGPE from the PGPE of the c-field method discussed earlier; in fact, the more recent formulations of the c-field method (SPGPE) are essentially equivalent to Stoof's theory (see Ref.~\cite{proukakis_jackson_08}).

In applying this general formalism, however, it is important to note that Stoof and coworkers make two further simplifications.  As in the ZNG theory, the thermal cloud excitations are treated within a simple self-consistent Hartree--Fock approximation.  However, more importantly (for the purposes of numerical implementation) the non-condensate component (thermal cloud) is assumed to be in {\it static} thermal equilibrium, described by the usual equilibrium Bose-Einstein distribution.  As noted earlier, this assumption is also standard in current applications of the c-field approach which include the coupling to the incoherent region (SPGPE).  However, explicit calculations using the ZNG coupled equations (see Chapter~12 of GNZ) have demonstrated that the inclusion of the thermal cloud dynamics is essential in the determination of the frequency and damping of condensate collective modes at higher temperatures.  The static thermal cloud approximation (see Chapter~8 of GNZ) may be useful as a starting point in understanding the effect of the thermal cloud, but it does not lead to quantitative predictions.

One might argue that the neglect of the dynamic HF mean field arising from the non-condensate atoms is a reasonable approximation since it is small in comparison to the condensate mean field.  However, we emphasise that the neglect of the thermal cloud HF mean-field $2gn'(\mathbf{r}, t)$ in the Stoof SGPE~\cite{stoof_99} is not a conceptual difference (as suggested in Ref.~\cite{proukakis_jackson_08}) with the generalised GPE of ZNG which includes it [see Eq.~(3.21) of GNZ], but is simply an {\it ad hoc} simplification useful for carrying out explicit calculations within the Stoof formalism.  By doing so, important physics is lost.  For one example, as discussed in Chapter~13 of GNZ, this thermal cloud HF mean-field is the agent responsible for the Landau damping of condensate collective modes.

\section*{Acknowledgements}
\emph{The original version of this manuscript was prepared to initiate a private electronic discussion between Allan Griffin on one side, and Tod Wright, Matthew Davis and Nick Proukakis on the other.  The goal of these correspondences was to understand each other's viewpoints on the utility of classical-field techniques for Bose gases as compared to the Beliaev broken symmetry approach.  Allan wrote the first draft of this manuscript with the intention that it might appear, alongside a response, as a chapter of the edited volume \emph{Quantum Gases: Finite Temperature and Non-Equilibrium Dynamics}, N.~P. Proukakis, S.~A. Gardiner, M.~J. Davis, and M.~H. Szymanska, eds., Imperial College Press, London (in press).  Following Allan's passing, Eugene Zaremba kindly agreed to continue the discussion, and prepared the final version of this manuscript.} 

We thank Nick Proukakis for spirited exchanges on this topic and Matt Davis and Tod Wright for useful discussions. This research was funded by a Discovery Research Grant from NSERC of Canada.

\bibliographystyle{prsty}

\end{document}